\documentclass{elsart}

\usepackage{epsfig}

\newcommand{\kl}{\mbox{$K_{L}$}}
\newcommand{\ks}{\mbox{$K_{S}$}}

\newcommand{\pip}{\mbox{$\pi^{+}$}}
\newcommand{\pim}{\mbox{$\pi^{-}$}}

\newcommand{\mevcsq}{\mbox{MeV/$c^{2}$}}
\newcommand{\gevc}{\mbox{GeV/$c$}}
\newcommand{\gevcsq}{\mbox{GeV/$c^{2}$}}

\newcommand{\listhead}[1]{\huge \noindent #1 \begin{itemize}}

\usepackage{amssymb}

\begin{document}

\begin{frontmatter}



\title{Electron Identification in Belle}


\author{K.~Hanagaki\thanksref{label1}\corauthref{cor1}}
\ead{kazu@bmail.kek.jp}
\corauth[cor1]{Corresponding author. ph +81-298-79-6080,
               fax +81-298-64-5340.}
\author{H.~Kakuno\thanksref{label2}},
\author{H.~Ikeda\thanksref{label3}},
\author{T.~Iijima\thanksref{label3}}, and
\author{T.~Tsukamoto\thanksref{label3}}

\address[label1]{Department of Physics, Princeton University,
Princeton, NJ, 08544}

\address[label2]{Department of Physics, Tokyo Institute of Technology,
Tokyo 152-8551}

\address[label3]{High Energy Accelerator Organization (KEK),
Tsukuba 305-0801}

\begin{abstract}

We report on electron identification methods and their performance in
the Belle experiment at the KEK-B asymmetric B-Factory $e^{+} e^{-}$ 
storage ring.   
Electrons are selected using a likelihood approach that takes
information  from the electromagnetic calorimeter, the central drift
chamber, and the silica aerogel  Cherenkov counters as
input.
We achieve an electron identification efficiency of $(92.4 \pm 0.4)\%$
with a $\pi^{\pm}$ fake rate of $(0.25 \pm 0.02)\%$ for the momentum range
between 1.0~GeV/$c$ and 3.0~GeV/$c$ in laboratory frame.

\end{abstract}


\end{frontmatter}


\section{Introduction}

The primary goal of the Belle experiment, which is conducted at the
KEK-B asymmetric $e^{+}$ (3.5~GeV) $e^{-}$ (8.0~GeV) collider,
is to measure the  $CP$ violation parameter
$\sin 2\phi_{1}$ in the neutral $B$ meson system~\cite{ref:sin2phi1}.
This measurement is done by studying the time evolution of events 
of the form
$e^+ e^- \rightarrow \Upsilon(4S) \rightarrow B^0 \overline{B^0}$
where one $B^0$ decays to a $CP$ eigenstate (for example
$B^0 \rightarrow J/\Psi K_S$) and the other decays to a final  state
that reveals its ``flavor''~\cite{ref:cp}.
In this context ``flavor'' refers to whether the non-$CP$-eigenstate
decay particle (the ``tagging'' $B^0$) is a $B^0$ or a
$\overline {B^0}$.
Electron identification (EID), which includes both $e^+$ and $e^-$,
plays an important role in this measurement since one of the most
reliable methods of determining a $B^0$  meson's flavor is to observe
the sign of the charged lepton that emerges when it undergoes
semi-leptonic decay.

In addition, EID is useful for reconstructing the large class of $CP$
eigenstates that involve final-state $J/\Psi$ particles, since these
are detected via the decay $J/\Psi \rightarrow \ell^+ \ell^-$ where
electron and muon pairs occur in equal numbers.  

Moreover, EID is crucial for analyses of
$b \rightarrow c$ (or $u$) $e^{-} \bar{\nu}$ semileptonic decays,
which enable us to extract $V_{cb}$ or $V_{ub}$~\cite{ref:CKM}.
For these purposes, electrons with the laboratory momenta
above 1~GeV/$c$ and below 3~GeV/$c$ must be identified with high
efficiency and high purity.

This paper describes the method used to discriminate between electrons
and other charged particles (mostly pions).  Its performance will also
be given.  In Section~2 , a brief description of the Belle detector is
given.  In Section~3 the details of the EID technique are presented
and in Sections~4 and 5 the performance results in terms of efficiency 
and fake rate are reported. The last section concludes this study.


\section{The Detector}

\begin{figure}[htbp]
  \begin{center}
    \leavevmode
    \epsfig{file=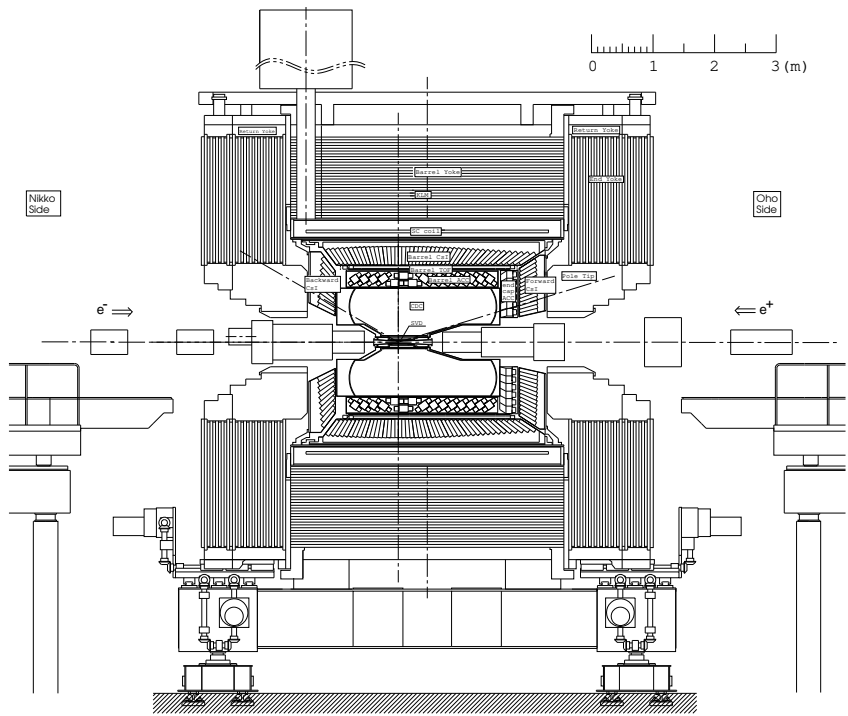,width=100mm,angle=270}
    \caption{
             Plan view of the Belle detector.
            }
    \label{fig:detector}
  \end{center}
\end{figure}

The Belle detector, which is shown in Fig.~\ref{fig:detector},
consists of a silicon vertexing detector for precise vertex
finding, a central drift chamber (CDC) for detecting charged
particles, arrays of silica aerogel Cherenkov counters
(ACC) for particle identification and time of flight counters
for particle identification, an electromagnetic calorimeter
(ECL) for photon and electron detection, a system of resistive plate
chambers interspersed with the magnetic return yoke for $\mu/\kl$
detection (KLM), and an extreme forward calorimeter for online
luminosity measurement.

A  1.5-T magnetic field is excited by a superconducting solenoid
situated between the ECL and the KLM.  
Additional details of the detector can be found in
\cite{ref:belle_nim}.
The sub-detectors relevant to EID are briefly described below.

The CDC is a cylindrical chamber with the inner radius of 77~mm, outer
radius of 880~mm, and a length of 2400~mm.
The geometrical coverage in $\theta$ is
$17^{\circ} < \theta < 150^{\circ}$~\footnote{
The forward direction is defined as $e^{-}$ beam direction.
}.
In addition to the tracking, the $dE/dx$ measurement that is made
by the CDC plays a significant role in EID.
The chamber volume comprises 8400 approximately rectangular drift
cells, arranged in 50 cylindrical layers.
The outer 47 layers are used for the $dE/dx$ measurement.
The gas filling  in the CDC is 50\% $He$ and 50\% $C_{2}H_{6}$
at atmospheric pressure.

The ECL, which is the primary detector used in EID, is an array of
8736 tower shaped CsI (T$l$) crystals that roughly project to the 
beam-beam interaction point (IP).  The ECL  consists of a barrel
and an endcap part.  
Each crystal is 30~cm (16.2~$X_{0}$) in depth  and roughly
$5 \times 5$~$\rm cm^{2}$ in cross section.   The 
ECL covers a polar angle of $12^{\circ} < \theta <
155^{\circ}$ in the lab frame.
Scintillation light from each crystal is read out by a pair of
silicon PIN photodiodes of cross section of
$1 \times 2$~$\rm cm^{2}$ mounted at the rear end of the crystal.
The ECL is designed to provide excellent energy resolution for
particles that induce electromagnetic showers.  Moreover, 
the center of gravity of the energy depositions from such 
showers  can be used to accurately determine the particles' 
position of incidence.   The energy and position
resolutions are given by
\[
(\frac{\sigma_E}{E})^2 = (\frac{0.066(\%)}{E})^2
                     + (\frac{0.81(\%)}{E^{1/4}})^2
                     + (1.34(\%))^2
\]
and
\[
  \sigma_{\rm pos}~({\rm mm}) = 0.27
                     + \frac{3.4}{E^{1/2}}
                     + \frac{1.8}{E^{1/4}}
~~~(E~{\rm in}~{\rm GeV}).
\]

The ACC consists of 1188 light-tight modules holding blocks 
of silica aerogel that are viewed by photomultipliers (PMT's).
The array, which consists of a barrel part and an endcap part like the
ECL, covers the polar angle range between $17^{\circ}$ and
$127^{\circ}$.
The refractive indices of the blocks vary from 1.010 to 1.030
depending on the polar angle because a high momentum particle
tends to direct forward region and vice-versa,
due to the asymmetric beam energy.
Although the main task of ACC is $K/\pi$ separation,
it is also useful for EID below the $\pi$ meson threshold at
$\sim 1$~\gevc. 


\section{Method of Electron Identification}

In order to distinguish  electrons from  hadrons (or muons), two
approaches  are used in our EID scheme.
The first exploits the major difference in the electromagnetic
showers induced by the electrons and the hadronic showers induced
by the pions and other hadrons.  In particular, we make use of
the significant difference in energy deposition and shower shape
between the two types of particles.  
The second makes use of the difference in velocity for electrons and
hadrons of the same momentum.  This difference, which
is largest at low momentum,   enables us to discriminate between
electrons and hadrons through $dE/dx$ measurements and through
observation of the light yield in the ACC array.
Information from the two approaches is combined into a single
variable using a likelihood method.

\subsection{Principle -Likelihood Method-}

The basic goal here is to combine EID information from various
discriminants, many of which when used in isolation provide only
modest separation between electrons and other particles, into a single
quantity that provides optimal discrimination power.  
To do this, we calculate the likelihood for each discriminant
based on probability density functions (PDFs) prepared beforehand.
For each discriminant, the electron likelihood ($L_{e}$), and
the non-electron likelihood ($L_{\bar{e}}$), are separately
calculated.
Next each likelihood is combined using
\[
L_{eid} = \frac {\prod_{i=1}^{n} L_{e}(i) }
     {\prod_{i=1}^{n} L_{e}(i) + \prod_{i=1}^{n} L_{\bar{e}}(i) } \: ,
\]
where $i$ runs over each discriminant.
Since we do not apply a correction to compensate for possible
correlations between the discriminating variables, the output of EID,
$L_{eid}$, is not a probability, but nonetheless is still useful for
discriminating between electrons and other particles. 

Note that a particular $L_{e}$ or $L_{\bar{e}}$ will return a
value of 0.5 if there is no pertinent information to distinguish
between electrons and hadrons. 
For example, ECL information is of no value in cases where the
particle does not hit the ECL, for example for tracks having
transverse momenta low enough that they curl up inside of the
CDC.

\begin{figure}[htbp]
  \begin{center}
    \leavevmode
    \epsfig{file=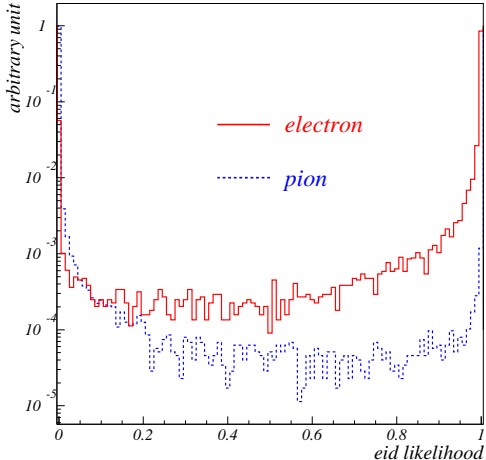,height=70mm,width=70mm}
    \caption{Likelihood ratio.
             Electrons and pions are denoted by solid
             broken histograms, respectively.
            }
    \label{fig:prob}
  \end{center}
\end{figure}

Figure~\ref{fig:prob} shows the distribution of $L_{eid}$ using all
the discriminants described below.
Both electrons and charged pions are shown.
As can be seen, significant e/$\pi$ separation is achieved
with this variable.

\subsection{The Discriminants}

We use the following five discriminants in the EID:
\begin{enumerate}
\item Matching between the position of the charged track extrapolated 
      to the ECL and the energy cluster position measured by the ECL.
\item Ratio of the energy measured by the ECL, $E$, and the charged track
      momentum measured by the CDC, $p$.
\item Transverse shower shape at the ECL.
\item $dE/dx$ in the CDC.
\item Light yield in the ACC.
\end{enumerate}
Below we discuss each discriminant in detail.
In each case data are used to show the detector performance for
electrons and hadrons.
For electrons, a sample of radiative Bhabha events are used, while for
hadron backgrounds $\ks \rightarrow \pip \pim$ decays in
hadronic events are used. 
Since the first item, matching between CDC tracks and ECL clusters,
is needed to derive the $E/p$ and the shower shape at the ECL,
we describe the matching first.
We then go through the remaining discriminants: ECL, CDC, and ACC
in order.

\subsubsection{Position Matching of the ECL Cluster to the Charged
               Track} 
\label{sec:matching}

The position matching between charged tracks and ECL clusters 
contributes to the EID in two ways.  First, it helps to
discriminate between electrons and hadrons directly, 
since the position resolution for electron showers is 
considerably smaller than that for hadronic showers.  Second, 
proper position matching is essential to derive the correct ratio for
$E/p$, which gives the largest discriminating power between
electrons and hadrons. 

\vspace*{1em}
To obtain reliable matching between the extrapolated track and the
center of gravity of the electron's shower, it is necessary to
ensure that the track is extrapolated to the appropriate depth
in the calorimeter.  This is particularly important at low
momenta, where the track curvature can be significant.
Thus we use a momentum dependent extrapolation depth,
which is derived from Monte Carlo (MC) simulation~\cite{ref:GEANT}.
The extrapolation direction is determined using the momentum
vector of the incident charged particle at the front surface 
of the ECL.

Defining the difference between the cluster position and the position
of an electron track extrapolated to the ECL in $\phi$ (azimuth) as
$\Delta \phi$, and in $\theta$ (polar angle) as $\Delta \theta$, we
define a matching $\chi^{2}$ as 
\[
  \chi^{2} \equiv \left(\frac{\Delta \phi}
                     {\sigma_{\Delta \phi}}\right)^{2}
        + \left(\frac{\Delta \theta}
                     {\sigma_{\Delta \theta}}\right)^{2} \; ,
\]
where $\sigma$ is the width obtained by fitting the $\Delta \phi$ and
$\Delta \theta$ distributions for electrons with Gaussian functions.
For each charged track, the cluster giving the minimum $\chi^{2}$ is
taken for the matched cluster, and is used to calculate $E/p$ and to
derive the transverse shower shape.  If no cluster with a matching
$\chi^{2}$ less than 50 is found, the track is considered to have
no associated ECL cluster.

\begin{figure}[htbp]
  \begin{minipage}[t]{65mm}
  \begin{center}
    \leavevmode
    \epsfig{file=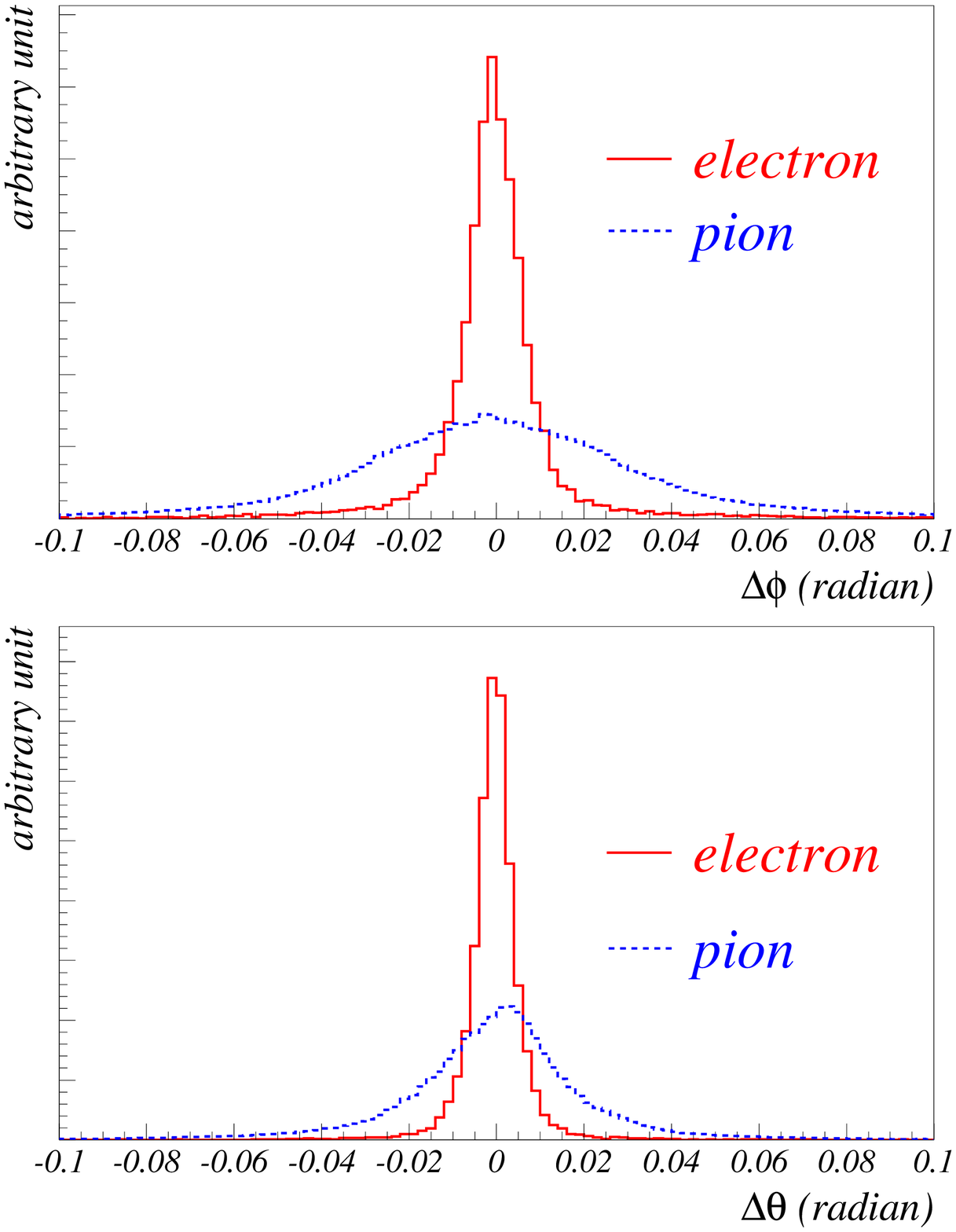,height=65mm,width=65mm}
    \caption{$\Delta \phi$ (top) and $\Delta \theta$ (bottom).
             The solid (broken) line shows the electron
             (charged pion) case.}
    \label{fig:dphidthe}
  \end{center}
  \end{minipage}
  \hfill
  \begin{minipage}[t]{65mm}
  \begin{center}
    \leavevmode
    \epsfig{file=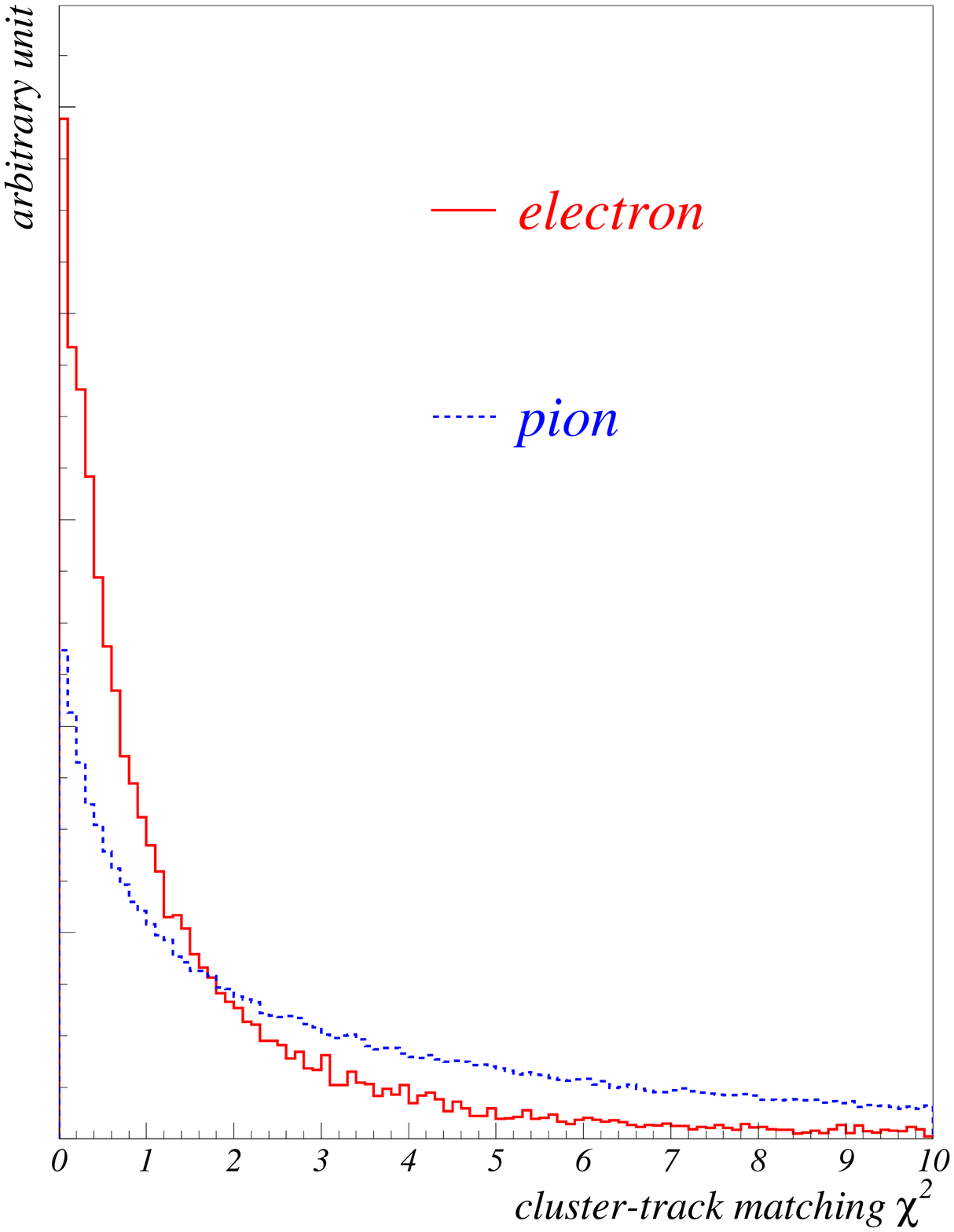,height=65mm,width=65mm}
    \caption{Cluster-track matching $\chi^{2}$.
             The solid (broken) line shows the electron
             (charged pion) case.
             }
    \label{fig:mat}
  \end{center}
  \end{minipage}
\end{figure}

\vspace*{1em}
Figure~\ref{fig:dphidthe} shows $\Delta \phi$ (top)
and $\Delta \theta$ (bottom) for electrons and pions.
As can be seen, electron clusters exhibit better track matching
than the pion clusters do.  
The position matching $\chi^{2}$ distributions for electrons and pions
are shown in Fig.~\ref{fig:mat}.
This $\chi^{2}$ distribution is used as one of the discriminants.

\subsubsection{$E/p$}

Since electrons in the energy range of interest have
negligible mass, we have $E = \sqrt{p^2 + m^2} \simeq p$
and we expect that the ratio $E/p = 1$ within measurement
errors.  For pions and other hadrons $E/p$ is typically smaller
than one, and, more importantly, the energy deposition for
hadronic showers is highly variable. 

\begin{figure}[htbp]
  \begin{minipage}[t]{65mm}
  \begin{center}
    \leavevmode
    \epsfig{file=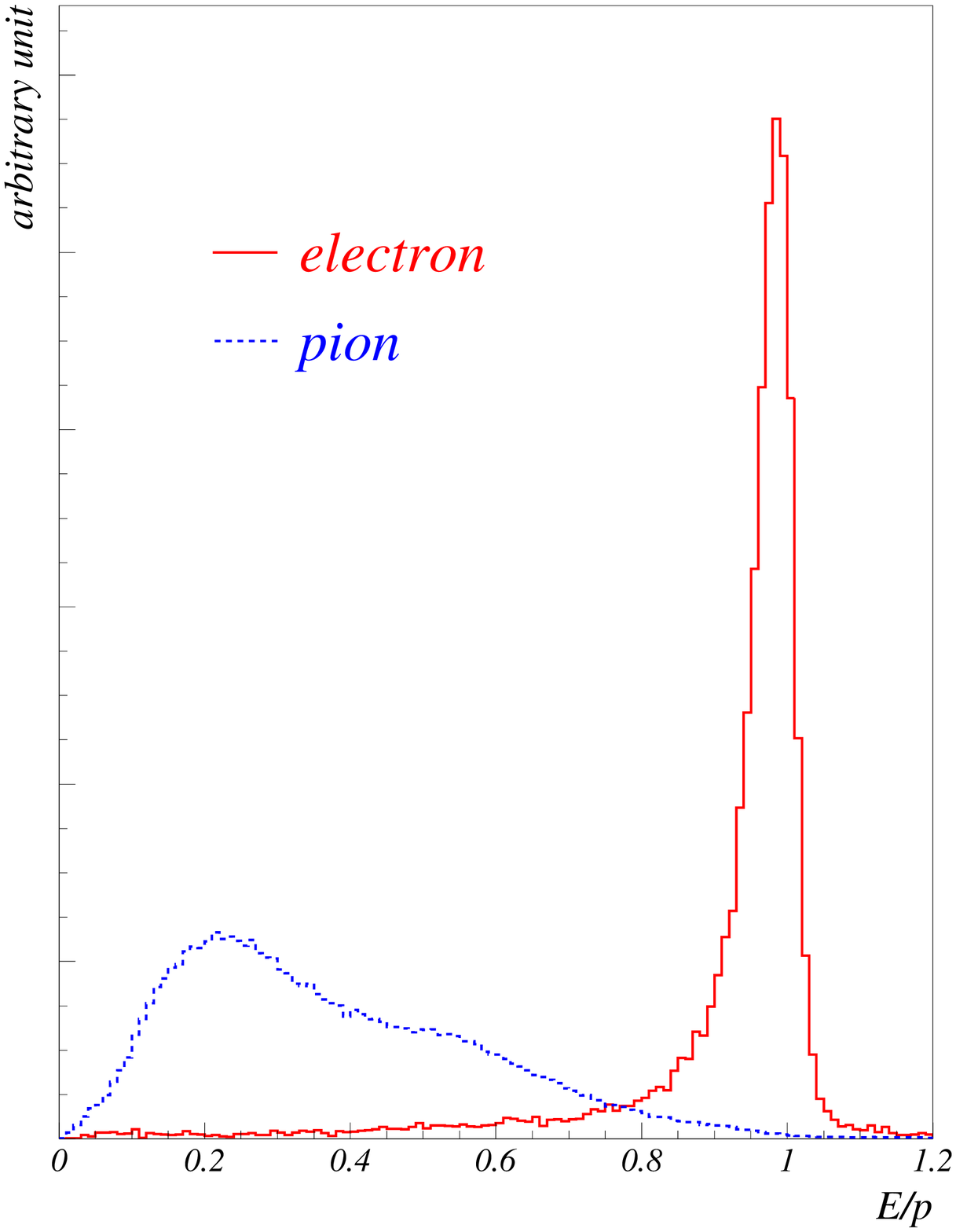,height=65mm,width=65mm}
    \caption{$E/p$ for electrons (solid)
             and charged pions (broken).
            }
    \label{fig:eop}
  \end{center}
  \end{minipage}
  \hfill
  \begin{minipage}[t]{65mm}
  \begin{center}
    \leavevmode
    \epsfig{file=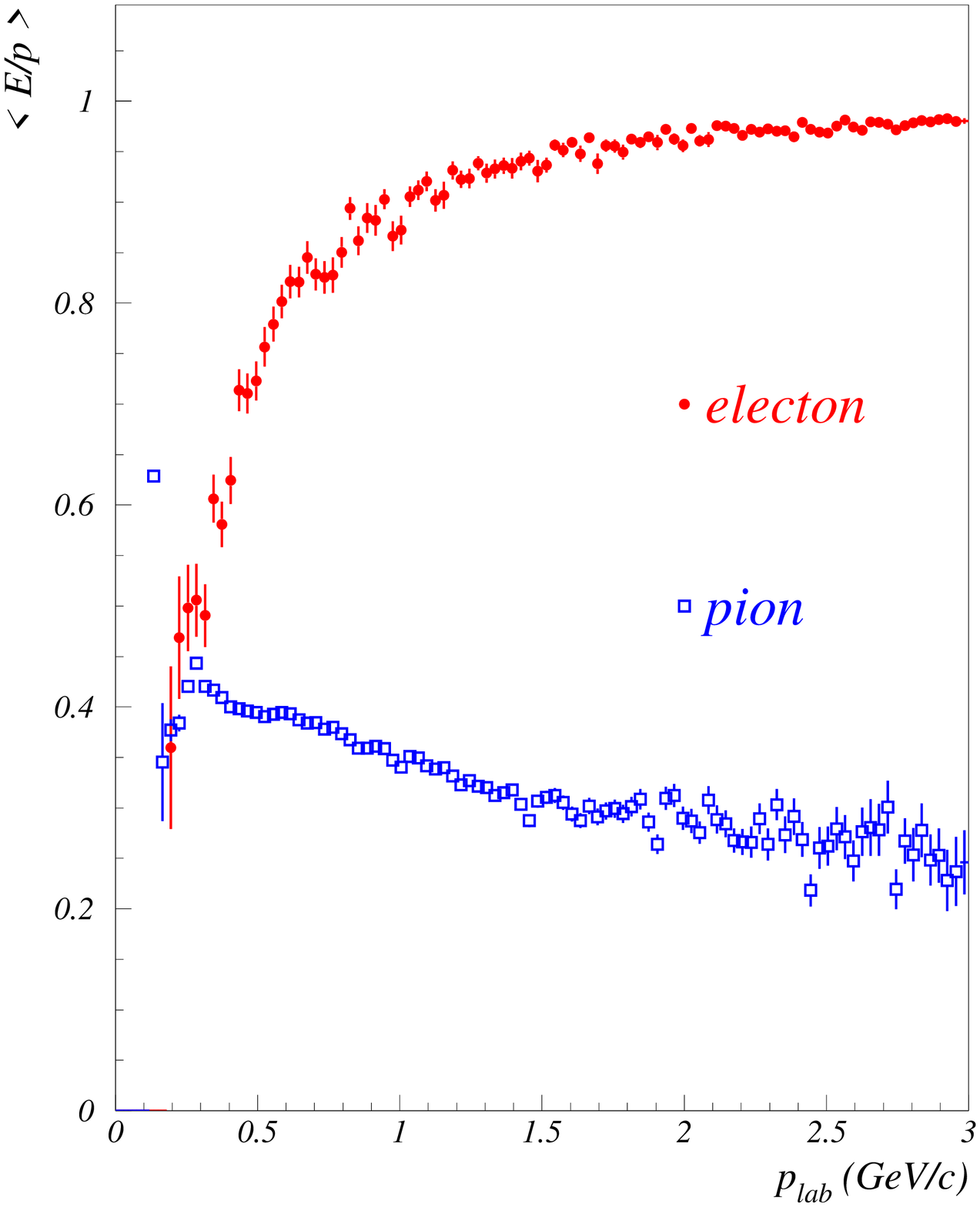,height=65mm,width=65mm}
    \caption{Momentum dependence of $E/p$.
             Electrons are represented by filled circles,
             and charged pions by open squares.
             }
    \label{fig:eop_plab}
  \end{center}
  \end{minipage}
\end{figure}

Figure~\ref{fig:eop} shows $E/p$ distribution for electrons in 
the momentum region $0.5 < p_{\rm lab} <3.0$~\gevc~\footnote{
Note that the momentum spectrum for electrons in radiative Bhabha 
events is completely different from that for electrons in hadronic
events. 
In particular, high-momentum electrons dominate radiative Bhabha
events. Thus, one should not assume that the $E/p$ distribution in
Fig.~\ref{fig:eop} is the same as the $E/p$ distribution 
for electrons in hadronic events. }.
Also shown is $E/p$ for charged pions.
As can be seen, $E/p$ gives a large discriminating power between
electrons and hadrons.
For example, a simple cut of $E/p > 0.8$ would give the
efficiency of 76.1\% for electrons, and 3.4\% for pions.

The lower-side tail in the electron $E/p$ distribution comes from
the interactions of electrons with material in front of the ECL.
Figure~\ref{fig:eop_plab} shows the momentum dependence of $E/p$.
The lower the momentum, the larger the fraction of events that have 
$E/p$ different from unity as a result of interactions. 
On the other hand, pions have a lower $E/p$ at higher the momenta,
so that $E/p$ is an efficient parameter in EID for high momentum
particles.

\subsubsection{Shower Shape}

Electromagnetic and hadronic showers have different shapes in
both the transverse and the longitudinal directions.
To evaluate the shower shapes in the transverse direction
quantitatively, we use the quantity $E9/E25$, which is defined 
as the ratio of energy summed in a 3~$\times$~3 array of crystals
surrounding the crystal at the center of the shower 
to that of a sum of a 5~$\times$~5 array of crystals centered
on the same crystal.  

\begin{figure}[htbp]
  \begin{center}
    \leavevmode
    \epsfig{file=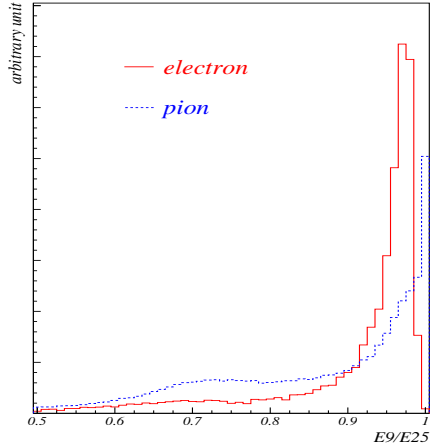,height=65mm,width=65mm}
    \caption{$E9/E25$ for electrons and pions denoted by solid
              broken histograms, respectively.}
    \label{fig:e9e25}
  \end{center}
\end{figure}

Figure~\ref{fig:e9e25} shows $E9/E25$ for electrons and charged
pions. Electrons exhibit a peak at around $E/p=0.95$, with a 
relatively small low-side tail, while pions have more events 
in the lower $E9/E25$ range.  
This is attributed to the faster evolution (measured in terms
of material depth) of electromagnetic showers
relative to that for hadronic showers.  
The events in the region of the pion distribution near unity arise
from minimum ionizing energy deposit in a single ECL block.

\subsubsection{$dE/dx$}

\begin{figure}[htbp]
  \begin{minipage}[t]{65mm}
  \begin{center}
    \leavevmode
    \epsfig{file=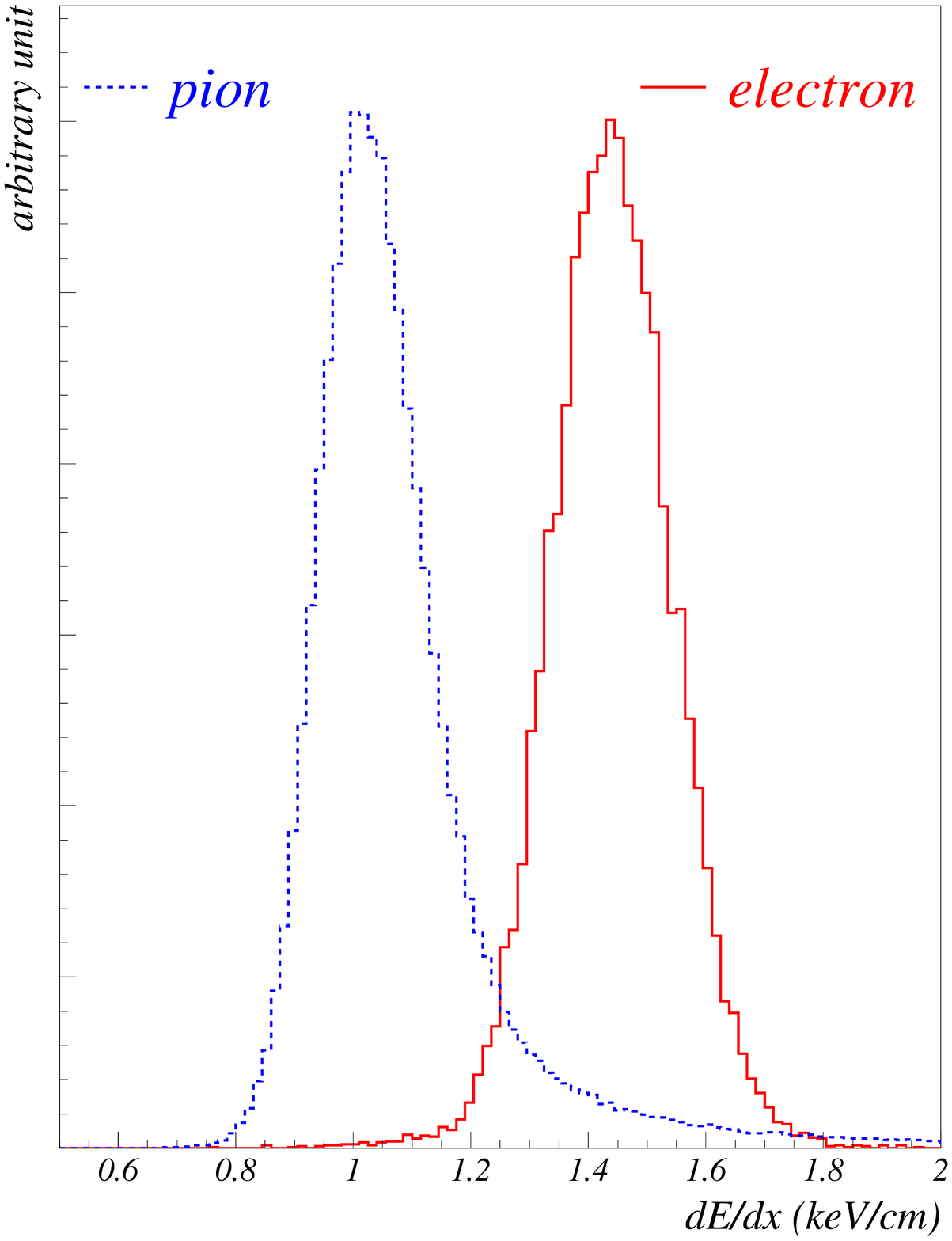,height=65mm,width=65mm}
    \caption{$dE/dx$ for electrons and pions.
            The solid histogram represents electrons, and
            the broken histogram represents pions.
            }
    \label{fig:dedx}
  \end{center}
  \end{minipage}
  \hfill
  \begin{minipage}[t]{65mm}
  \begin{center}
    \leavevmode
    \epsfig{file=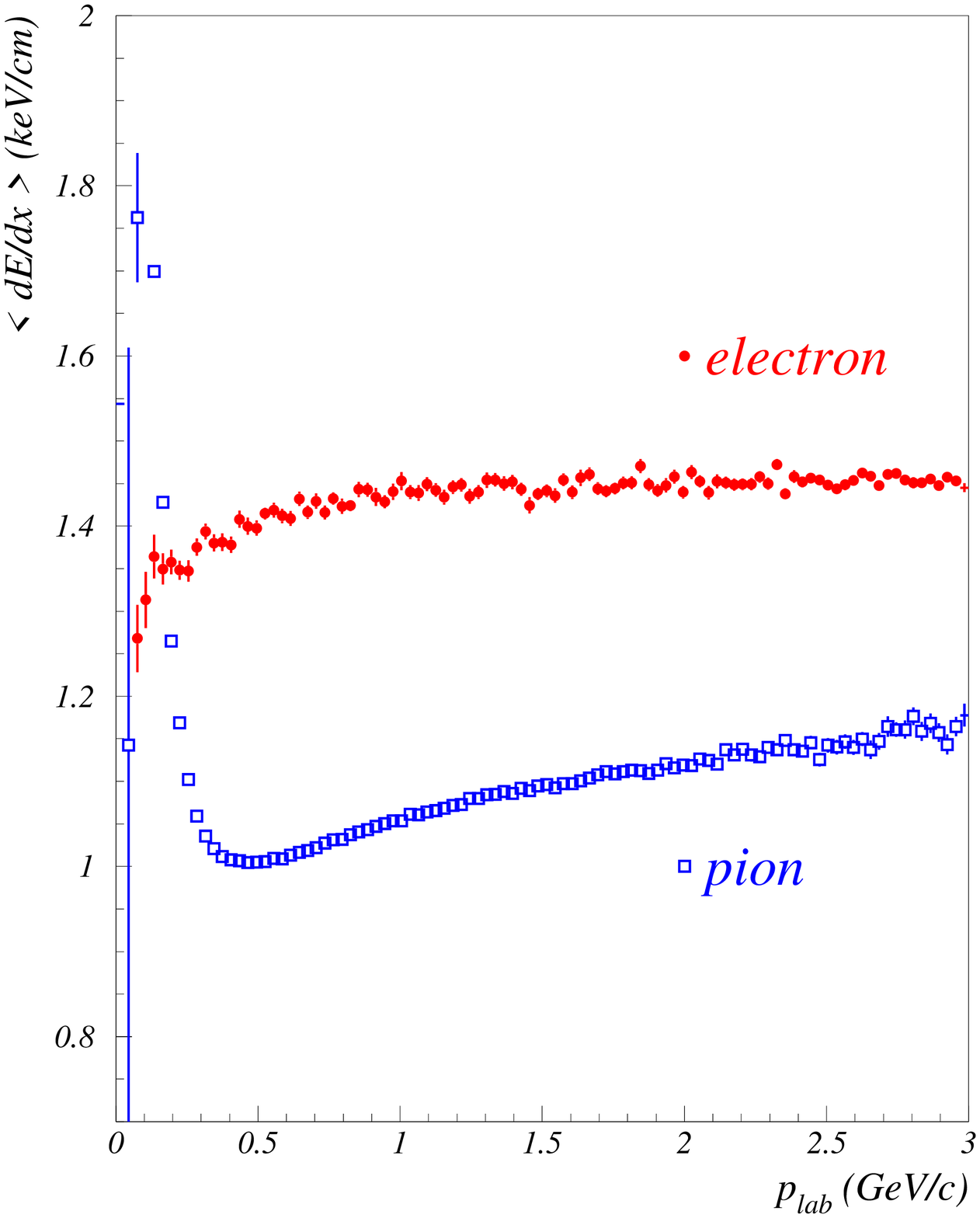,height=65mm,width=65mm}
    \caption{Momentum dependence of $dE/dx$.
             Electrons are represented by filled circles,
             and pions by open squares.
             }
    \label{fig:dedx_plab}
  \end{center}
  \end{minipage}
\end{figure}

The amount of ionization created by a particle as it travels through
a gas filled volume is proportional to its rate of energy loss,
$dE/dx$, which exhibits a well-known $\beta^{-2}$ dependence.
Although the statistical fluctuations for any single $dE/dx$
measurement are typically quite large, and exhibit a pronounced
high-side tail, an accurate determination of $dE/dx$ can be made by
averaging several measurements.
By excluding  the highest 20\% of the individual measurements
from the average, the effects of upward fluctuations can be
suppressed.
Figure~\ref{fig:dedx} shows the resulting $dE/dx$ distributions
for electrons and for pions.  The resolution for pions is 7.8\%.

The averaged $dE/dx$ as a function of momentum is shown in
Fig.~\ref{fig:dedx_plab}.
As can be seen, the momentum range of interest is 
almost fully covered by $dE/dx$ in terms of e/$\pi$ separation,
although the separating power is higher in the lower momentum
region.  This approach  is complementary to the $E/p$ approach.

The likelihood for $dE/dx$ is calculated from the measured $dE/dx$
($(dE/dx)_{\rm meas}$), expected $dE/dx$ ($(dE/dx)_{\rm exp}$), and
the expected resolution ($\sigma_{dE/dx}$), assuming the PDF
to be a Gaussian.
The mean of the Gaussian is deduced from the Bethe-Bloch equation as a
function of velocity based on an arbitrary particle hypothesis.
The expected resolution is determined from test beam results.
Given a $\chi^{2}$ of 
\[
\chi^{2} = ( \frac{ (dE/dx)_{\rm meas} - (dE/dx)_{\rm exp} }
                  { \sigma_{dE/dx} } )^{2} \; ,
\]
the likelihood probability, $L$, is expressed as
\[
L = \frac{ e^{-\frac{1}{2}\chi^{2}} } { \sqrt{2\pi} \sigma_{dE/dx} }
   \; .
\]

\subsubsection{Light Yield in the ACC}

In the ACC, the Cherenkov threshold for electrons is just
a few MeV, while that for pions is 0.5 to 1.0~\gevc, depending on the
refractive index.  Therefore, the ACC is able to distinguish between 
electrons and hadrons in the momentum region below $\sim 1.0$~\gevc.
A likelihood is calculated from
the light yield of  the aerogel radiator (actually by the number of
photoelectrons from the  PMT) using PDFs calculated from MC 
distributions for 20 different velocity ranges.

\subsection{PDFs for Discriminants Related to the ECL}
\label{sec:PDF}

PDFs for $E/p$, $E9/E25$, and the position matching $\chi^{2}$ are
created in a manner similar to the one described below. Two sets of
PDFs are prepared, one for data, and another for MC.
For producing electron PDFs in real data  analysis, we use
real $e^{+}e^{-} \rightarrow e^{+}e^{-}e^{+}e^{-}$ events.

The PDFs for non-electrons are generated using hadronic MC events.
This is done for both data PDFs and MC PDFs.  The MC sample used
for this purpose consists of generic $B-\bar{B}$ 
decays and continuum events in a 1:3 ratio.   Random-trigger
events  from the experiment are overlaid on the MC events to simulate
accidental backgrounds.  
The shapes of the PDFs are parametrized by fitting the distributions of
each discriminant to shape functions selected by trial and error.
A summary of the selected shape functions appears in
Table~\ref{table:fit}.
\begin{table}[htbp]
 \begin{center}
  \caption{Shape of PDFs.}
  \label{table:fit}
  \begin{tabular}{l|l}
   discriminant	&function modeling the shape \\
   \hline
   $E/p$ for $e^{\pm}$	&$f(x) = (1-E) \times A \times
      \exp [-\frac{1}{2} (\frac{x-B}{\sigma})^{2}]$ \\
    & $+ E \times A \times \exp [-\frac{1}{2} (\frac{x-B}{F})^{2}]$ \\
    &where the $\sigma$ is $C+D(x-B)$ if $x<B$, and is $C$ if $x>B$.\\
   $E/p$ for $\pi^{\pm}$	&triple Gaussian + linear \\
   $E9/E25$ for $e^{\pm}$	&$g(x) = A \times \exp [-\frac{1}{2}
                       (\frac{x-B}{C + D(x-B)})^{2}]$ \\
   $E9/E25$ for $\pi^{\pm}$	&$h(x) = g(x)$ + Gaussian \\
   matching $\chi^{2}$ for $e^{\pm}$	&exponential + linear \\
   matching $\chi^{2}$ for $\pi^{\pm}$	&exponential + linear \\
  \end{tabular}
 \end{center}
\end{table}

To take into account the momentum and polar angle dependence of the
PDFs, the fits to the PDF functions are carried out in ten momentum
bins and six polar angle bins (total of 60 bins in all).
The bins are divided into
\begin{itemize}
\item 0.0-0.6, 0.6-0.7, 0.7-0.82, 0.82-0.95, 0.95-1.1, 1.1-1.27,
      1.27-1.47, 1.47-1.7, 1.7-2.05, and $>$2.05~GeV/$c$ in lab. frame
      momentum;
\item $< 33.6^{\circ}$ (forward endcap region),
      $33.6^{\circ}$ - $45.9^{\circ}$,
      $45.9^{\circ}$ - $70.6^{\circ}$,
      $70.6^{\circ}$ - $109.4^{\circ}$, 
      $109.4^{\circ}$ - $132.5^{\circ}$, and
      $>132.5^{\circ}$ (backward endcap region) in polar angle.
      The polar angle bins correspond to the boundaries of six different
      groups of CsI crystals in the ECL.
\end{itemize}


\section{Efficiency}

Two different types of event samples are used to study the
efficiency of the EID.  The first are QED events which provide
a clean source of electrons over a wide momentum range.
These serve as a good calibration sample.  However, to take
into account possible efficiency degradations for electrons
in hadronic events, where there are more charged
tracks, more ECL energy clusters, and more accidental hits in all
detectors, we also studied the EID efficiency in a hadronic
environment.

We here define EID efficiency as
\[
\rm efficiency = \frac{ Number\; of\; tracks\; identified\;
                        as\; an\; electron }
                      { Number\; of\; electron\; tracks \;
                        found\; by\; tracking } \; ,
\]
where the threshold is applied at 0.5 on $L_{\rm eid}$.
Note that the tracking efficiency is not included here.
Unless otherwise mentioned, the charged tracks used in this
study are required to come from the IP.~\footnote{
Defining $dr$ ($dz$) as the closest approach to the IP in $r$-$\phi$
($r$-$z$) plane, $|dr| < 0.5$~cm and $|dz| < 1.5$~cm are required.
}

\subsection{Efficiency in QED Events}
\label{sec:eff_in_qed}

\begin{figure}[htbp]
  \begin{minipage}[t]{65mm}
  \begin{center}
    \leavevmode
    \epsfig{file=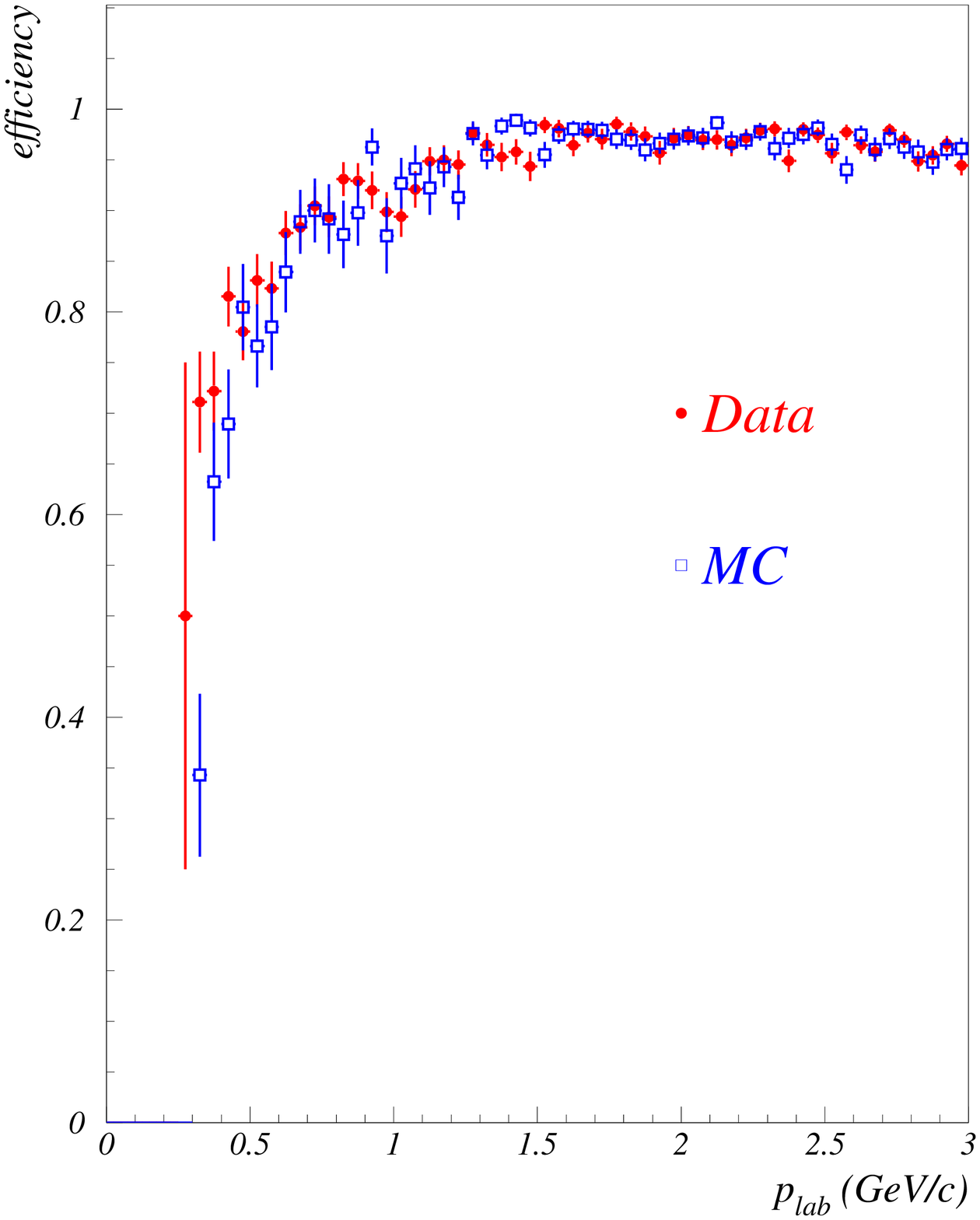,height=65mm,width=65mm}
    \caption{Efficiency in radiative Bhabha events as a function of
             lab momentum. 
             Data are represented by filled circles,
             and MC simulation by open squares.
             }
    \label{fig:eff_mom}
  \end{center}
  \end{minipage}
  \hfill
  \begin{minipage}[t]{65mm}
  \begin{center}
    \leavevmode
      \epsfig{file=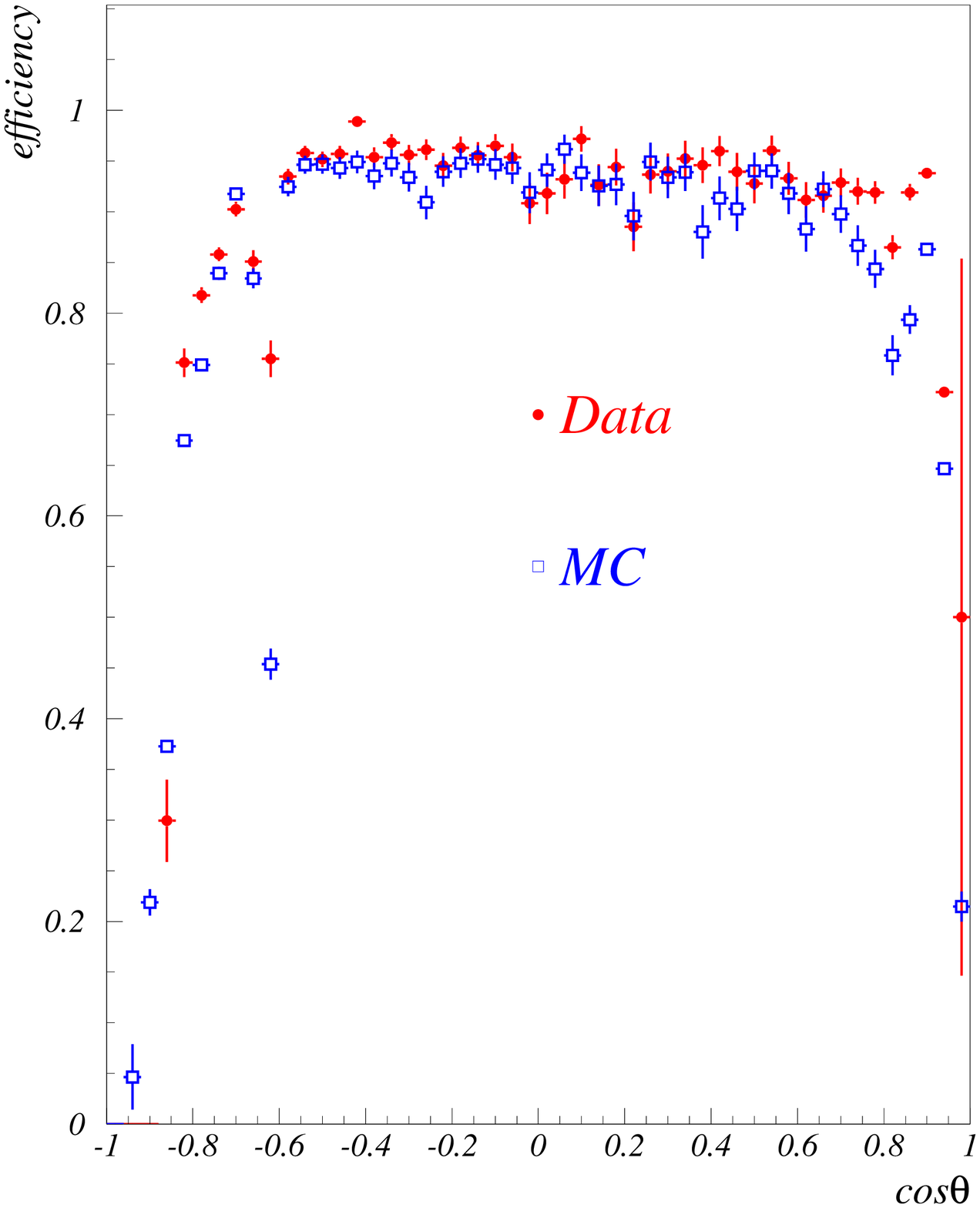,height=65mm,width=65mm}
      \caption{Efficiency in radiative Bhabha events as a function of
               $\cos \theta$ of a track.
               Data are represented by filled circles,
               and MC simulation by open squares.
              }
      \label{fig:eff_theta}
  \end{center}
  \end{minipage}
\end{figure}

The EID efficiency in radiative Bhabha events~\footnote{
We consider tau-pair and hadronic events as the background sources.
MC study shows that these background sources are negligible.
}
is measured and compared with
MC expectations as shown in Fig.~\ref{fig:eff_mom}.
The polar angle here is restricted to be
$-0.57 < \cos\theta < 0.82$.  This eliminates
the very forward and very backward regions, where the EID efficiency
is difficult to evaluate owing to tracking instabilities brought
on by large beam-induced backgrounds.   
The measured efficiency in the momentum range above 1~GeV/$c$ is
over 90\% and is in good agreement with MC prediction.
This momentum region includes most of the range populated by
electrons from $J/\psi \rightarrow e^{+}e^{-}$ decays as well as
most of the electrons used in flavor tagging.  

The polar angle dependence of the efficiency is shown in
Fig.~\ref{fig:eff_theta}.
The momentum range of the electrons is
$0.5 < p_{\rm lab}< 2.5$~GeV/$c$.
Although the agreement between data and MC is good, the efficiency
itself in the endcap region is much lower than that in the barrel.
This is because more material exists in the endcap region, which
degrades both the ECL's resolution and the CDC's tracking
performance.
The low efficiency seen at $\cos \theta \sim -0.6$ is due to a
gap between the barrel part and endcap part of the ECL.

\subsection{Efficiency in Hadronic Events}

The EID efficiency in hadronic events is studied using
1) single-electron MC tracks embedded in real hadronic events, and
2) $J/\psi (\rightarrow e^{+}e^{-})$ inclusive events.

\subsubsection{Single $e^{\pm}$ MC Embedded in Hadronic Events}
\label{sec:eff_embed}

A comparison between the efficiency obtained for isolated single-track
electrons and for electron tracks embedded in
hadronic data events is shown in Table~\ref{table:eff_hadronic}.
Since the polar angle distributions for the hadronic
MC and the single electron MC tracks are different, only the 
$-0.57 < \cos \theta < 0.82$ region is used for this
comparison.
\begin{table}[htbp]
 \begin{center}
  \caption{
           Efficiency (\%) in the region of
           $-0.57 < \cos\theta < 0.82$.
          }
  \label{table:eff_hadronic}
  \begin{tabular}{c|c|c|c}
   $p_{\rm lab}$(GeV)	&single MC
	&single MC in hadronic events	&generic hadronic MC	\\
   \hline
   0.00-0.25 &$41.0 \pm 5.4$ &$42.0 \pm 5.9$ &$51.5 \pm 1.2$\\
   0.25-0.50 &$72.1 \pm 1.2$ &$68.2 \pm 1.4$ &$71.0 \pm 1.0$\\
   0.50-0.75 &$83.5 \pm 0.9$ &$81.7 \pm 1.0$ &$82.4 \pm 0.9$\\
   0.75-1.00 &$91.2 \pm 0.7$ &$89.2 \pm 0.8$ &$90.4 \pm 0.8$\\
   1.00-1.25 &$94.4 \pm 0.6$ &$90.8 \pm 0.7$ &$93.0 \pm 0.8$\\
   1.25-1.50 &$96.4 \pm 0.5$ &$93.4 \pm 0.6$ &$95.1 \pm 0.8$\\
   1.50-1.75 &$98.0 \pm 0.4$ &$94.8 \pm 0.6$ &$96.0 \pm 0.8$\\
   1.75-2.00 &$97.5 \pm 0.4$ &$96.2 \pm 0.5$ &$95.5 \pm 1.0$\\
   2.00-2.25 &$98.1 \pm 0.3$ &$95.8 \pm 0.5$ &$96.7 \pm 1.1$\\
   2.25-2.50 &$98.7 \pm 0.3$ &$95.7 \pm 0.6$ &$95.3 \pm 1.9$\\
  \end{tabular}
 \end{center}
\end{table}

\begin{figure}[htbp]
  \begin{center}
    \leavevmode
    \epsfig{file=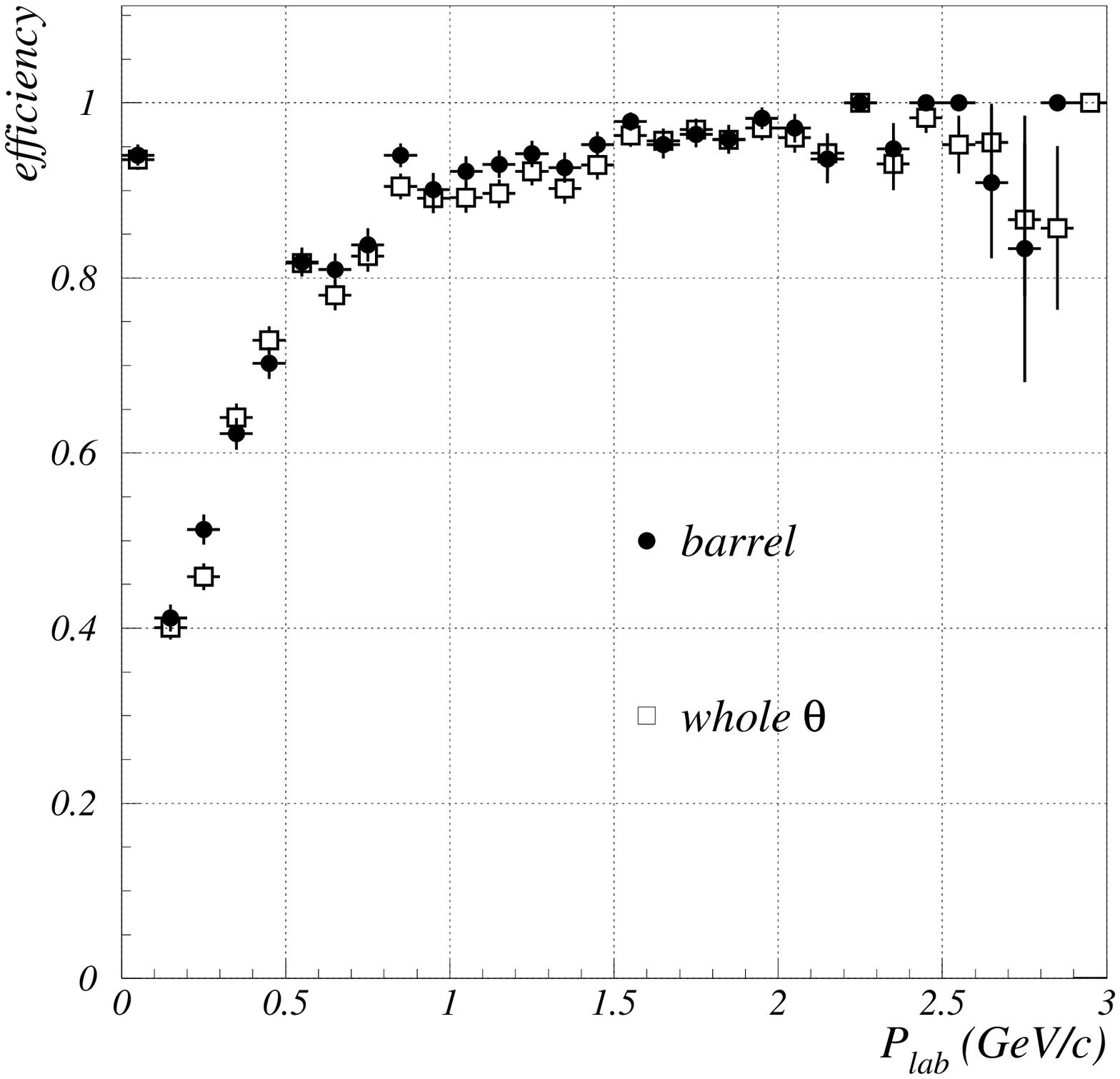,width=65mm}
    \epsfig{file=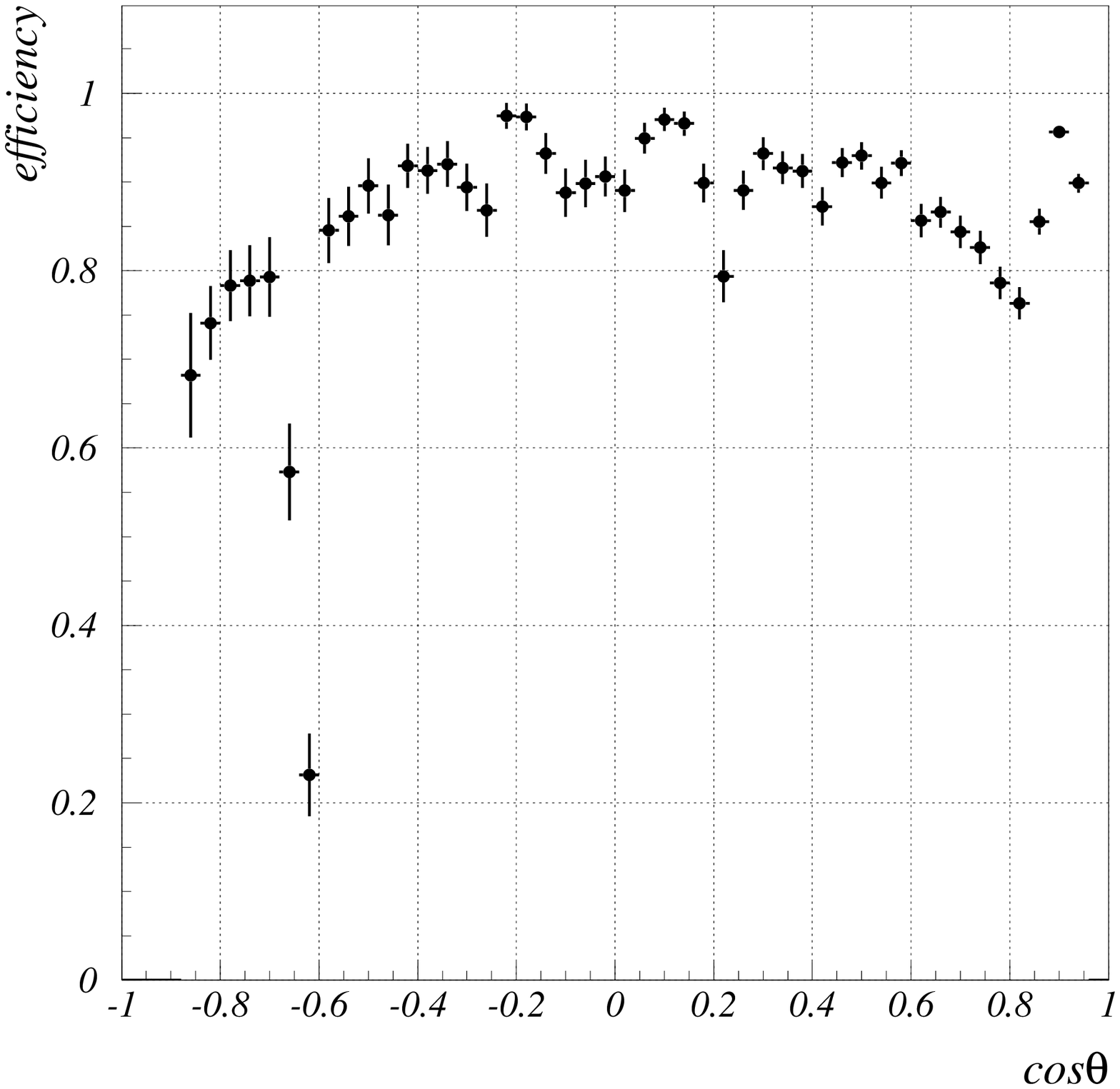,width=65mm}
    \caption{
             Efficiency in the generic hadronic MC.
             The left plot shows the momentum dependence of the EID
             efficiency.
             The efficiency for $-0.57 < \cos \theta < 0.82$
             is illustrated by filled circles, and that for the full 
             region by open squares.
             The right plot shows the polar angle dependence for the
             momentum greater than 0.5~\gevc\ in the lab frame.
            }
    \label{fig:eff_generic}
  \end{center}
\end{figure}

The efficiency for MC-generated single-track electrons having  
momenta greater than 0.5~GeV/$c$ drops by 2.5\% when these
tracks are embedded into hadronic events from the data 
sample.   This degradation is significant for the  $dE/dx$ and
the cluster-track matching efficiencies.
No statistically significant degradation is seen in $E/p$ or 
$E9/E25$.  The good agreement between the single-track MC electrons
embedded in hadronic events and the generic hadronic MC implies that 
the efficiency drop due to the hadronic environment is well reproduced 
in our generic-hadron MC.
We will therefore refer to the efficiency estimated by the generic
hadronic MC as the EID efficiency since the momentum and angular
distributions imitate  hadronic data.
This reference efficiency is shown in Fig.~\ref{fig:eff_generic}.
For momentum region $1.0~(0.5) < p_{\rm lab} < 3.0$~\gevc\ and for the
whole polar angle range, the efficiency is
$(92.4 \pm 0.4)\%$ ($(87.3 \pm 0.3)\%$).
One note here is that there is a strong correlation between lab frame
momentum and polar angle because of the asymmetric beam energy.
For example the forward endcap region seems to have higher efficiency
than that of barrel from Fig.~\ref{fig:eff_generic}, however, this
effect is caused by the harder momentum spectrum in the forward
region.
The barrel region has higher efficiency for the same momentum.

In the next subsections, we further examine the efficiency in hadronic
events by comparing data and MC.

\subsubsection{$J/\psi (\rightarrow e^{+}e^{-})$ Inclusive Events}

\begin{figure}[htbp]
  \begin{center}
    \leavevmode
    \epsfig{file=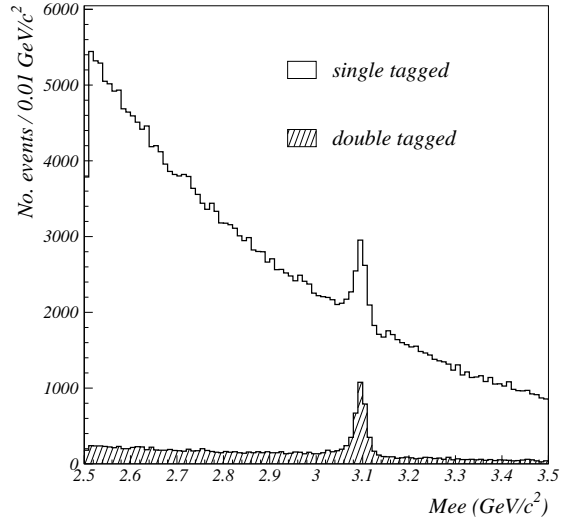,width=3.0in}
    \caption{
             $M_{ee}$ distributions.
             The open (hatched) histogram shows
             $M_{ee}$ with one (both) track(s) identified as an
             electron.
            }
    \label{fig:mee_exp7}
  \end{center}
\end{figure}

One can also obtain the EID efficiency or inefficiency by comparing
the $J/\psi (\rightarrow e^{+}e^{-}) X$ yield for the cases
one or two electrons are tagged, or the difference of the two cases.
To estimate the signal yield, we use the $M_{ee}$ distribution as
shown in Fig.~\ref{fig:mee_exp7}.
The sample is selected using
the standard hadronic event selection in Belle:
\begin{itemize}
\item The number of good tracks is required to be $\geq 3$, where
      a good track is defined to be one with $|dr|<2.0$~cm,
      $|dz|<4.0$~cm, and $p_{t}>0.1$~GeV/$c$.  
\item The primary vertex position has to be $<1.5$~cm in radius and 
      within $\pm 3.5$~cm of the IP in $z$.
\item The sum of charged-track momentum and ECL cluster energy in the
      $\Upsilon(4S)$ rest frame is $\geq 20\%$ of the center-of-mass 
      energy ($\equiv W$).
\item $|\sum p_{z}^{*}| < 50\%$ of $W$, where 
      $p_{z}^{*}$ is $z$ component of momentum for either charged
      tracks or ECL cluster energies in the center-of-mass frame.  
\item The sum of ECL cluster energies is between 2.5\% and 90\% of $W$.
\end{itemize}
In addition the ratio of the 2nd Fox-Wolfram moment~\cite{ref:FW} to
the 0th moment, R2, is required to be below 0.5 to suppress continuum
events.
Because there are many QED related backgrounds such as radiative
Bhabha events in which the photon is converted to an $e^{+}e^{-}$
pair, we require that the selected events have at least one electron
in the  barrel region and we raise the good-track requirement to $\geq
5$.
A requirement of $L_{eid}>0.5$ is used for tagging.
The electron momentum spectrum for the surviving events turns on 
at around 0.7~GeV/$c$ and extends up to $\sim 3.5$~GeV/$c$ with a
peak at about 1.6~GeV/$c$ in the laboratory  frame.

To model the signal shape, we first fit the distribution of double 
tagged $M_{ee}$ events with a 
``crystal-ball function''~\footnote{
A lineshape function first
introduced by the Crystal Ball collaboration, which provides a good
empirical match to the $J/\Psi \rightarrow e^+ e^-$ peak.}
for the signal and exponential for the background.
The same procedure is carried out for the sample of single-tagged
events, and that of single-tagged events subtracted by double-tagged
events.
In the fitting of these $M_{ee}$ spectra, the signal
peak position and width are fixed to the values extracted from the
double-tagged $M_{ee}$ spectrum, with only the signal yield being
allowed to float freely. 

Using the procedure above, the signal yield for the mass range between
2.5 and 3.5~\gevcsq\ is estimated to be $4107.9 \pm 112.0$ for single
tagging, and $540.2 \pm 131.6$ for the difference between single and
double tagged events.
Since the signal yield with single tagging (difference between single
tagging and double tagging) is proportional to $1-\bar{\epsilon}^{2}$
($2\bar{\epsilon}(1 - \bar{\epsilon}$)), where
the $\bar{\epsilon}$ denotes inefficiency, an inefficiency
$(6.2 \pm 1.4)\%$ is deduced.
This is consistent with the inefficiency of $(5.6 \pm 0.1)\%$ that is
predicted by the MC.

\subsection{Charge Dependence}
\label{sec:chg_dep}

\begin{figure}[htbp]
  \begin{center}
    \leavevmode
    \epsfig{file=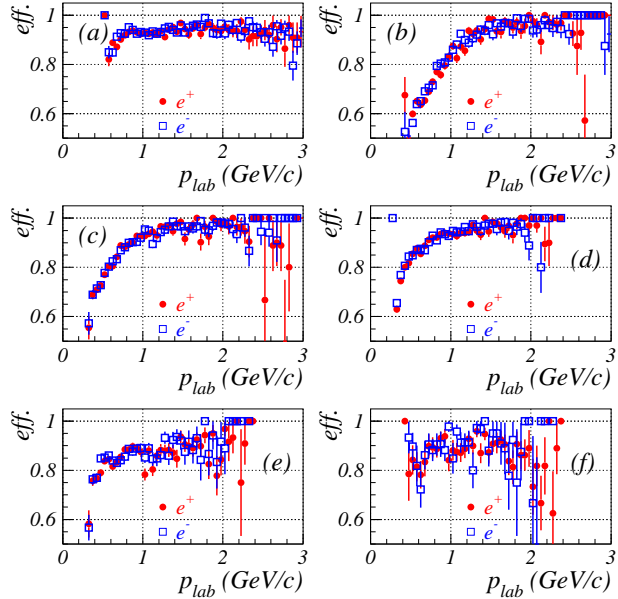,width=90mm}
    \caption{Efficiency in $e^{+}e^{-} \rightarrow
             e^{+}e^{-}e^{+}e^{-}$ events as a function of
	     momentum in lab. frame.
             The efficiency for positrons is denoted by filled
             circles, and that for electrons by open squares.
             Each plot corresponds to a different polar angle.
	     The (a) is for the forward endcap, (b) through (e) for
             the barrel, and (f) for the backward endcap.
             The division is the same as in the PDFs for likelihood
             using $E/p$ etc..
            }
    \label{fig:eff_chg}
  \end{center}
\end{figure}

Since a charge asymmetry in the efficiency would cause a bias in the
flavor tagging of initial $B$ states, it is important to examine this
possibility  closely. 
To obtain good statistics we used a sample of 
$e^{+}e^{-} \rightarrow e^{+}e^{-}e^{+}e^{-}$ events. 
Owing to the asymmetric beam energies, the polar angular 
distributions of electrons and positrons differ, so the 
efficiencies were compared for individual bins in momentum
and angle.
Figure~\ref{fig:eff_chg} shows the efficiency for  the
$e^{+}e^{-} \rightarrow e^{+}e^{-}e^{+}e^{-}$ sample divided into 
six different polar angle regions as a function of laboratory
momentum.
The polar-angle division corresponds to the six groups of PDFs
described in Section~\ref{sec:PDF}. 
The efficiencies for electrons and positrons are separately
illustrated.
No statistically significant difference in efficiency for the barrel
region between electrons and positrons is observed.
For the endcap region, the charge dependence still exists even with
this polar-angle division because of the different angular-momentum
correlation between electrons and positrons.
From these data one can extract 90\% confidence interval of
$-0.002 < A_{chg} < 0.002$ on the charge asymmetry for the momentum
range between 0.5~\gevc\ and 2.5~\gevc\ in the barrel region.


\section{Fake rate}
\label{sec:fake}

In addition to the efficiency, the fake rate is another
important factor in EID performance.   We define the fake rate as:
\[
\rm fake \; rate = \frac{ \#\; of \;non\; e^{\pm}\; tracks\;
                        found\; by\; tracking\; with\; the\;
                        {\it L}_{eid} > 0.5  }
                      { \#\; of\; non\; e^{\pm}\; tracks\;
                        found\; by\; tracking }
\]

In this section, we discuss the fake rate for $\pi^{\pm}$.
The comparison between data and MC, and the charge dependence are
described.
We also mention the measurement of fake rate for $K^{\pm}$.

\subsection{Fake rate to $\pi^{\pm}$}
\label{sec:fake_to_pi}

We use inclusive $\ks \rightarrow \pi^{+} \pi^{-}$ decays as a source
of charged pions to measure the EID fake rate.
The $\ks \rightarrow \pi^{+} \pi^{-}$ decays are selected from
two oppositely charged tracks
with the following selection criteria:
\begin{itemize}
\item The Belle standard hadronic event selection.
\item The distance in $z$ between the two helices at the decay vertex
      has to be less than 1~cm.
\item The deflection angle, defined as the angle between the $K_{S}$
      momentum vector and the vector extrapolated from the IP to the
      $K_{S}$ decay vertex, must be less than $15^{\circ}$.
\item The impact parameter in the $x$-$y$ view is required to be
      greater than 1~mm. 
\item The invariant mass reconstructed from the two pion tracks,
      $M_{\pip \pim}$, satisfies
      $487.7 < M_{\pip \pim} < 507.7$~\mevcsq.
\item One track of the pair is identified as a charged pion by
      requiring $L_{eid}<0.01$ and vetoing protons using another
      particle identification package for hadrons in
      Belle~\cite{ref:belle_nim}.  (No requirement is placed on
      the pion used to test the EID routines.)
\end{itemize}

\begin{figure}[htbp]
  \begin{minipage}[t]{65mm}
  \begin{center}
    \leavevmode
    \epsfig{file=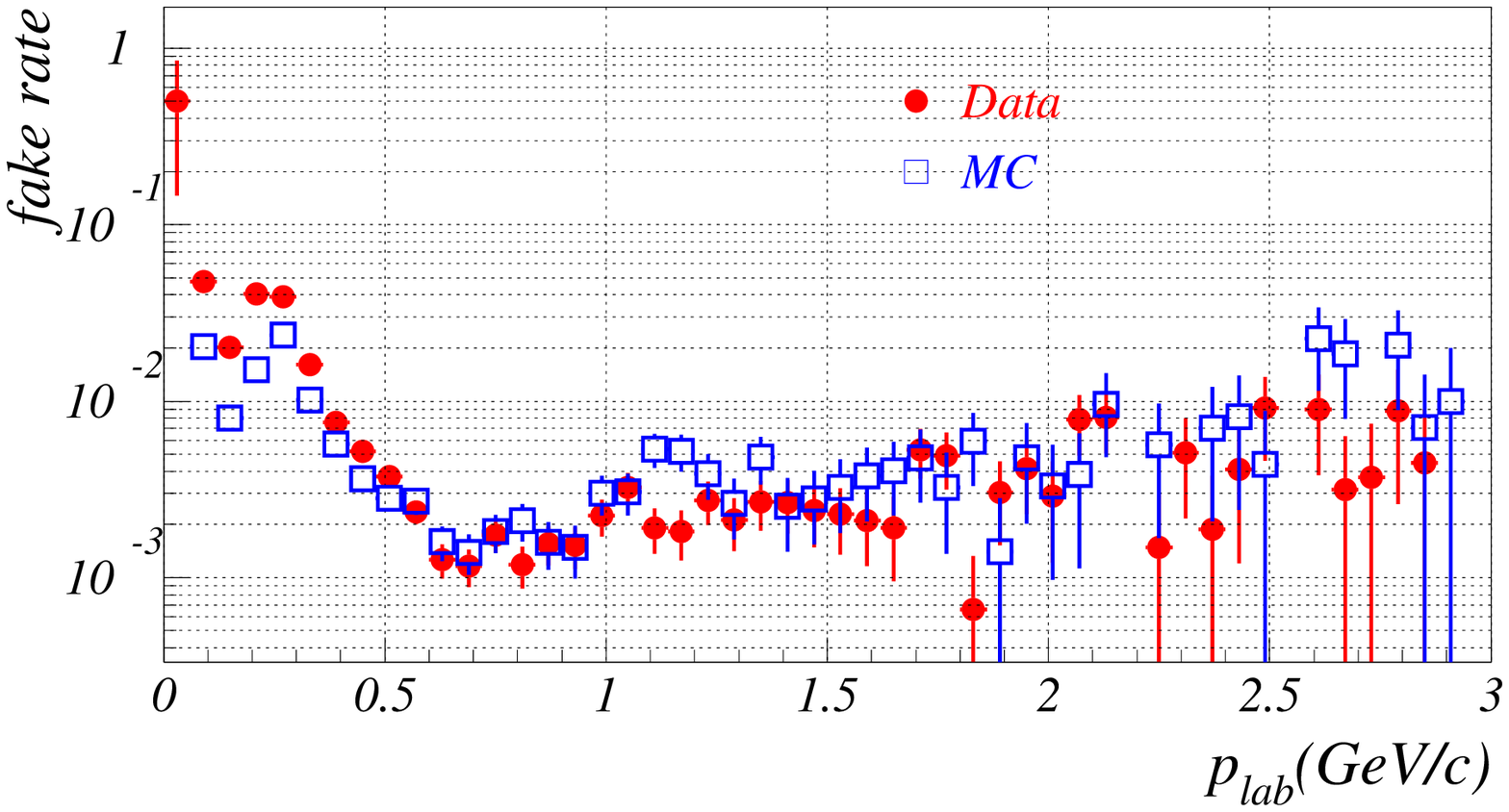,width=65mm}
    \caption{
             Fake rate for $\pi^{\pm}$ as a function of momentum.
             The data are shown as filled circles,
             and the MC as open squares.
            }
    \label{fig:fake_mom}
  \end{center}
  \end{minipage}
  \hfill
  \begin{minipage}[t]{65mm}
  \begin{center}
    \leavevmode
    \epsfig{file=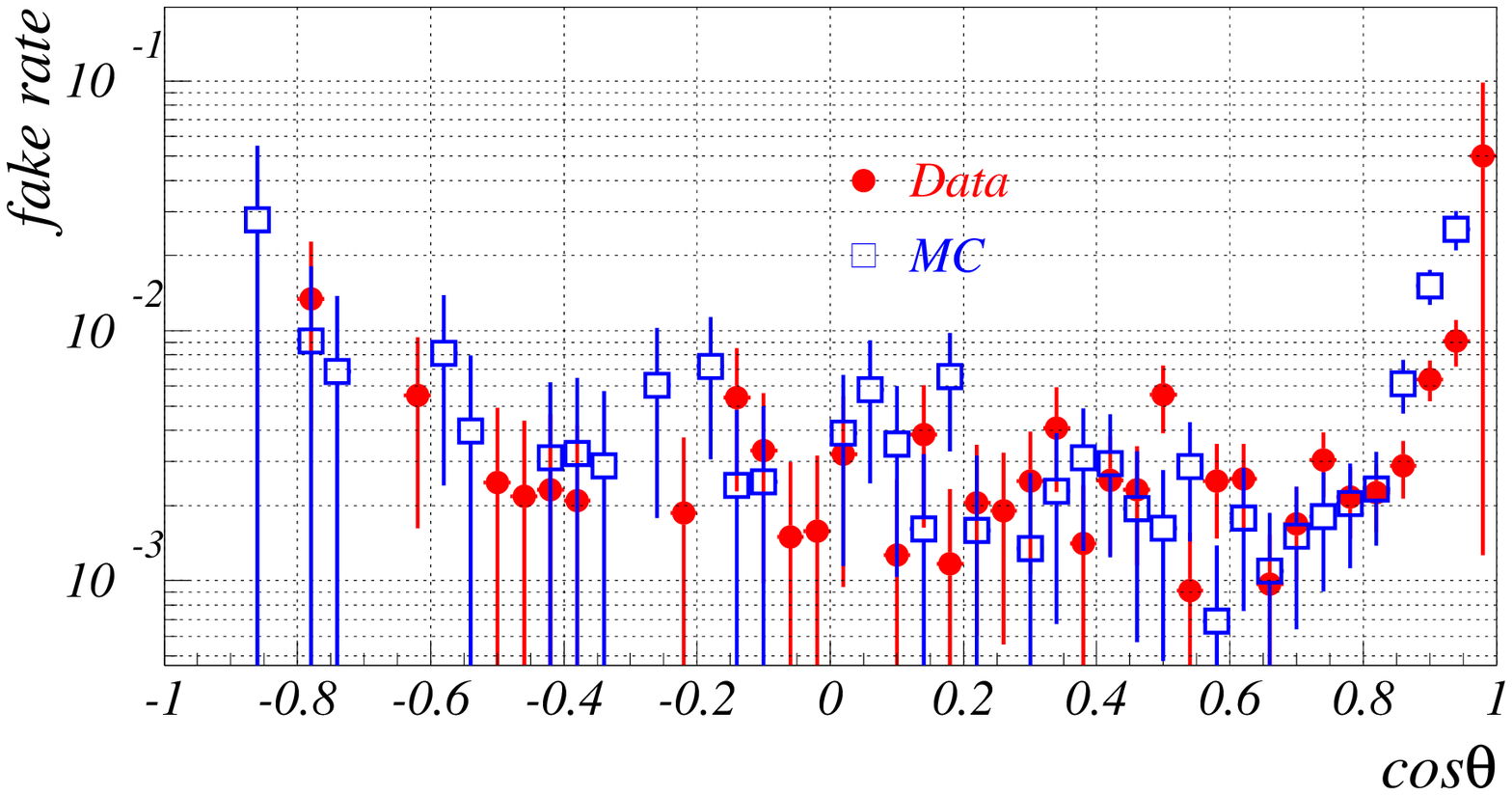,width=65mm}
    \caption{Fake rate for $\pi^{\pm}$ as a function of polar angle.
             The data are shown as filled circles,
             and the MC as open squares.
            }
    \label{fig:fake_theta}
  \end{center}
  \end{minipage}
\end{figure}

Figure~\ref{fig:fake_mom} shows the fake rate as a function of
laboratory momentum with a plot of the MC expectation superimposed.
Data and MC agree well for momenta above 0.5~\gevc.
The result obtained from this sample is also summarized in 
Table~\ref{table:fake_chgdep}.

The polar angle dependence is shown in
Fig.~\ref{fig:fake_theta}.
The overall agreement between data and MC is good, although the fake
rate in the data is slightly lower than MC expectation.
For example, the fake rate in the  momentum range between 0.5 and
3.0~\gevc\ is $(0.22 \pm 0.01)\%$ in data, and $(0.27 \pm 0.01)\%$ in
MC.
This disagreement mainly comes from an inconsistency in the forward
endcap region.
If the polar angle is restricted to the barrel region only 
($-0.57 < \cos\theta < 0.82$) the fake rates are 
$(0.16 \pm 0.01)\%$ for both data and MC.

\begin{table}[htbp]
 \begin{center}
  \caption{Fake rate (\%) to charged pions in real data.
          }
  \label{table:fake_chgdep}
  \begin{tabular}{c|c|c|c}
   $p_{\rm lab}$(GeV)	&$\pi^{\pm}$	&\pip\	&\pim\	\\
   \hline
   0.00-0.25 &$3.50 \pm 0.09$ 	&$3.57 \pm 0.13$   &$3.43 \pm 0.13$\\
   0.25-0.50 &$1.51 \pm 0.04$ 	&$1.50 \pm 0.05$   &$1.52 \pm 0.05$\\
   0.50-0.75 &$0.20 \pm 0.02$ 	&$0.24 \pm 0.03$   &$0.16 \pm 0.02$\\
   0.75-1.00 &$0.15 \pm 0.02$ 	&$0.19 \pm 0.03$   &$0.12 \pm 0.02$\\
   1.00-1.25 &$0.25 \pm 0.03$ 	&$0.26 \pm 0.04$   &$0.24 \pm 0.04$\\
   1.25-1.50 &$0.25 \pm 0.04$ 	&$0.30 \pm 0.06$   &$0.21 \pm 0.05$\\
   1.50-1.75 &$0.29 \pm 0.06$ 	&$0.35 \pm 0.09$   &$0.24 \pm 0.07$\\
   1.75-2.00 &$0.31 \pm 0.07$ 	&$0.46 \pm 0.13$   &$0.17 \pm 0.08$\\
   2.00-2.25 &$0.42 \pm 0.11$ 	&$0.28 \pm 0.12$   &$0.57 \pm 0.18$\\
   2.25-2.50 &$0.45 \pm 0.14$ 	&$0.18 \pm 0.12$   &$0.73 \pm 0.26$\\
   2.50-2.75 &$0.42 \pm 0.17$ 	&$0.42 \pm 0.24$   &$0.43 \pm 0.25$\\
   2.75-3.00 &$0.32 \pm 0.19$	&$0.00 \pm 0.00$   &$0.68 \pm 0.39$\\
  \end{tabular}
 \end{center}
\end{table}

\subsection{Charge Dependence in Fake Rate}

The charge dependence of the fake rate is also checked  using
$\ks \rightarrow \pi^{+} \pi^{-}$ events.
The fake rates for $\pi^{+}$ and $\pi^{-}$ are separately listed in
Table~\ref{table:fake_chgdep}.
As the table shows, the fake rate for \pip\ is
systematically higher than that for \pim\ for momentum below
2~GeV/$c$.
On the other hand, the fake rate for \pip\ is lower than that for
\pim\ above 2~GeV/$c$.
This is due to the different nuclear shower evolution
between \pip\ and \pim,
resulting in non-identical $E/p$ distributions.

\subsection{Fake Rate to $K^{\pm}$}
\label{sec:fake_k}

The fake rate for $K^{\pm}$ is examined using the decay chain
$D^{*+} \rightarrow D^{0} (\rightarrow K^{-} \pi^{+}) \pi^{+}_{s}$,
where the subscript of $\pi^{+}_{s}$ is used as a flag to signify
that the $\pi^{+}$ comes from $D^{*+}$ decay or $D^{0}$ decay.
This decay chain allows us to have a $K^{-}$ sample without any
particle identification because the charge of $\pi_{s}$ serves as a 
discriminant between $K^{-}$ and $\pi^{+}$ through its charge
correlation.

The strategy for evaluating the fake rate is to compare the signal
yield of $D^{0}$ with and without applying EID for the kaons.
The $D^{0}$ signal yield is measured by fitting the mass distribution
to the sum of a Gaussian for the signal and a linear function for the
background.
The selection criteria for the $D^{*}$ decay chain are:
\begin{itemize}
\item Defining $\theta$ as the angle between the $K^{-}$ flight
      direction in the $D^{0}$ rest frame and the $D^{0}$ flight
      direction in the $\Upsilon(4S)$ frame, $|\cos \theta| < 0.8$.
\item Defining $x$ as the $D^{*+}$ momentum in the $\Upsilon(4S)$ rest
      frame divided by the beam energy in that frame, $x$ is required
      to be $> 0.45$.
      This allows us to select continuum events,  simplifying comparison
      between data and MC.  
\item The reconstructed mass difference between the $D^{*+}$ and the $D^{0}$
      is within 1.5~MeV/$c^{2}$ from the nominal mass difference.
\end{itemize}

After the above selection, we have
16470$\pm$150.0 events without EID, and 71.4$\pm$18.5 events after
applying EID.
Taking the ratio, the fake rate for the kaons is measured to be
$(0.43 \pm 0.07)\%$
This can be compared to the MC prediction of $(0.21 \pm 0.05)\%$ for 
the fake rate.


\section{Conclusions}

Using a generic hadronic MC sample consisting of $B-\bar{B}$ and
continuum events in a 1:3 ratio with randomly triggered events
overlaid, the EID efficiency is estimated to be $(92.4 \pm 0.4)\%$ for
the momentum range $1.0 < p_{\rm lab} < 3.0$~GeV/$c$.
Using inclusive $K_{S} \rightarrow \pip \pim$ events,
the fake rate to charged pions is measured to be $(0.25 \pm 0.02)\%$
for $1.0 < p_{\rm lab} < 3.0$~GeV/$c$.

The EID efficiency expected from the generic hadronic MC is
consistent with that for single electrons in real hadronic data within
1\%.
Using $J/\psi \rightarrow e^{+} e^{-}$ events, the EID inefficiency is
verified to be consistent between data and MC
within a 1.4\% uncertainty.
The fake rate measured in data is consistent with the MC
expectation to 0.05\% in absolute value.
For a restricted theta range ($-0.57 < \cos\theta < 0.82$),
the fake rate in data agrees with MC expectations within 0.01\%.

%


\section*{Acknowledgments}

First we would like to thank the members of the Belle collaboration.
We acknowledge those who struggled to produce an accurate detector
calibration, which is essential to the electron identification.
The efforts of the ECL group were particularly helpful in this
regard.
Special thanks go to R.~Enomoto for organizing the EID group in
the early stages of this work.
We are grateful
Y.~Kwon and T.~Kim for starting up the original scheme
used in the likelihood calculation,
D.~Marlow for his very careful reading of the
manuscript and critical comments,
H.~Sagawa for fruitful discussion on the ECL,
and W.~Trischuk for his review of this manuscript.




\end{document}